\documentclass[usaAMS,usenatbib]{mn2e}

\usepackage[]{natbib}

\input epsf.sty
\usepackage{graphicx}
\usepackage{epsfig}
\usepackage{enumerate}
\usepackage{amssymb}
\usepackage{amsmath}
\usepackage{array}
\usepackage{balance}

\title{Spherical symmetry breaking in cold gravitational collapse of isolated systems}
\author[Tirawut Worrakitpoonpon]{Tirawut Worrakitpoonpon$^{1}$\thanks{E-mail: worraki@gmail.com} \\
$^{1}$Division of Physics, Faculty of Science and Technology, 
Rajamangala University of Technology Suvarnabhumi,  \\
Nonthaburi Campus, 7/1 Nonthaburi 1 Road, Nonthaburi 11000, Thailand}

\begin{document}

\pagerange{\pageref{firstpage}--\pageref{lastpage}} \pubyear{2014}

\maketitle

\label{firstpage}

\begin{abstract}
We study, using $N$-body simulation, the shape evolution in gravitational collapse of
cold uniform spherical system. The central interest is on how the deviation 
from spherical symmetry depends on particle number $N$. By revisit of the spherical 
collapse model, we hypothesize that the departure from spherical symmetry is regulated 
by the finite-$N$ density fluctuation. Following this assumption, the estimate of the 
flattening of relaxed structures is derived to be $N^{-1/3}$.
In numerical part, we find that the virialized states can be characterized by the 
core-halo structures and the flattenings of the cores fit reasonably well with the prediction. 
Moreover the results from large $N$ systems suggest the divergence of relaxation time 
to the final shapes with $N$.
We also find that the intrinsic shapes of the 
cores are considerably diverse as they vary from nearly spherical, prolate, oblate or 
completely triaxial in each realization. When $N$ increases, this variation is suppressed 
as the final shapes do not differ much from initial symmetry.
In addition, we observe the stable rotation of the virialized states. 
Further investigation reveals that the origin of this rotation is related in some way 
to the initial density fluctuation.
\end{abstract}



\maketitle


\section{Introduction} \label{intro}
Elliptical galaxies have been known for a long time and constituted one of the main 
categories of modern galaxy classification. 
Observationally it is clear that they are in some form of virialized
states with surface brightness suiting the well-known de Vaucouleurs profile 
(\citealp{de_vaucouleurs_1953}). In attempt to answer this, D. Lynden-Bell introduced the 
violent relaxation theory (see \citealp{lynden_bell_1967}) which
regarded the observed galaxies as meta-stable equilibria resulting from 
the collisionless relaxation. This evolution time-scale was notably short so that
the protogalaxies were able to attain the stationary states within a few crossing times.
Although the theory of Lynden-Bell was considered to be a milestone for understanding the 
collisionless dynamics of self-gravitating systems and, more generally, 
the systems governed by other long-range interactions,
it did not provide a satisfactory explanation of how the elliptical galaxies had obtained 
their shapes. As a consequence the question about the evolutionary background to those observed 
ellipticities then became the open research topic in galactic astrophysics and cosmology.

Current understanding suggests two distinctive formation schemes: merger of galaxies and 
violently collapsing protogalaxy. The former scenario has been suggested by 
\citet{toomre+toomre_1972, toomre_proc_1977} which involved at least two encountering 
galaxies under mutual gravitational interaction. It was shown that the final shape after the
merger could be elliptical. Numerous merging simulations producing galaxies similar to those from 
observations were reported (see, for example, \citealp{barnes+hernquist_1996} and references therein).
In the latter scenario, people were focusing on the isolated radially anisotropic system
under the gravitational force. It was long believed that such a system is highly unstable to the 
deformation of configuration.
The instability was first detected in the pioneering simulation by \citet{henon_1973},
who figured out that under a circumstance where system was dominated by radial motion, 
the spatial configuration of system was drastically deformed. 
Similar result was obtained later by \citet{polyachenko_1981}
in which, starting from the predominantly radial spherical system, an ellipsoidal core of structure 
was observed. These results gave rise to the so-called radial orbit instability (ROI), 
which was originally mentioned by \citet{antonov_book_1973} (see summary 
in English in \citealp{de_zeeuw_conf_1987}). 
For the isotropic system it was known that any distribution function was always 
stable as long as it was the decreasing function of energy per unit mass (see
\citealp{antonov_1961, doremus+feix+baumann_1971}).

Although the ROI theory did not provide any clue on
precise stability rule, many subsequent studies were able to determine, semi-analytically 
or numerically, the criteria at which shape deformation was triggered. 
Various diagnostics were then proposed. Studies led 
by \citet{merritt+aguilar_1985, palmer+papaloizou_1987, barnes+goodman+hut_1986}
showed that an anisotropic ratio calculated from twice the radial kinetic energy over 
the tangential kinetic energy was sufficient to evaluate the stability. Each group finished up 
with different numerical threshold values but they all shared the same conception that 
large anisotropic ratio increased the chance of shape deformation. On the other hand, groups of
\citet{min+choi_1989, cannizzo+hollister_1992, theis+spurzem_1999}
used an indicator as simple as the initial virial ratio. 
In the case of sufficiently low initial virial ratio, it was found that the 
violently collapsing phase occurred, which was analogous to the radially anisotropic system,
and the final shape was considerably changed. 
Later there was also the report of study using the combination of both parameters 
(see \citealp{barnes+lanzel+williams_2009}). 
In another aspect, it was proposed that the instability could be engendered by density 
contrast from the inhomogeneity of initial density profile (see \citealp{roy+perez_2004}). 
Later the same instability in dissipative systems was also examined by \citet{marechal+perez_2010}.

Question on intrinsic three-dimensional shape of elliptical galaxies is still  
unsolved. It was long believed that elliptical galaxy is oblate 
due to the flattening about the axis of rotation. Theoretical supports for oblate 
galaxy could be found in \citet{dehnen+gerhard_1994, robijn+de_zeeuw_1996} 
and also references therein. An objection was then 
made based on the observed rotational velocity of galaxies that led to 
another propositions for prolate galaxy (e.g. \citealp{binney_1978}) and triaxial galaxy
(e.g. \citealp{schwarzschild_1979, schwarzschild_1982, de_zeeuw_1985}). 
Numerical works of cold collapse experiments starting from various density profiles 
(see \citealp{aguilar+merritt_1990, cannizzo+hollister_1992, boily+athanassoula_2006} for example)
also produced different results. They found that the final intrinsic shapes were not unique 
but they depended strongly on the choices of initial conditions.
However, the investigation on this topic relies mainly on theoretical 
and numerical approaches due to the observational limitation.

In this paper we investigate the shape evolution in the cold collapse of
isolated self-gravitating system in spherical symmetry. 
We are interested in the aspects different from those in the past literatures: 
what is the role of finite-$N$ density fluctuation in the evolution from spherical 
symmetry and how can we describe, in the quantitative way, its 
influence on the final shapes. This paper is organized as follows. 
First in Section \ref{theory}, we revisit the dynamical 
model of cold gravitational collapse of spherical system
and derive the estimate of some variables in stationary states as a function of $N$
such as the flattening and specific angular momentum.
These predictions will be used in comparing with numerical results in further section.
In Section \ref{nume}, we 
introduce the initial condition and simulation parameters that we employ in this work. 
Next in Section \ref{results}, we present the numerical results, focusing mainly on the 
temporal evolution of systems and their properties in stationary states. 
Some of the results may be compared with theoretical predictions we derived earlier. 
Finally in Section \ref{conclusion}, we give the conclusion and discussion.


\section{Dynamics of cold gravitational collapse} \label{theory}

This section describes in detail our theoretical study of spherical cold collapse.
We first recall the dynamical model of this process in continuum limit. 
Then, we apply the density fluctuation and derive the estimate of some variables in 
stationary states as a function of $N$. Results from this section will be used in 
comparing with numerical results in Section \ref{results}.

\subsection{Shape evolution in system with finite $N$} \label{theory_evolution}
Consider a spherically symmetric system in continuum limit with uniform 
density $\rho_{0}$; the equation of motion of a thin spherical mass shell 
with radial position $r$ at time $t$ is given by
\begin{equation}
\frac{\partial^{2}r}{\partial t^{2}}=-\frac{GM_{r}}{r^{2}} \label{eq_motion}
\end{equation}
where $M_{r}=4\pi \int_{0}^{r}\rho_{0}r^{2}dr$ is the total mass inside $r$. 
Defining the scale factor $R=r/r_{0}$, 
the equation of motion can be rewritten in terms of $R$ as
\begin{equation}
\frac{\partial^{2} R}{\partial t^{2}}=-\frac{GM_{r}}{r_{0}^{3}R^{2}}. \label{eq_motion_r}
\end{equation}
For cold initial state, the integration of equation (\ref{eq_motion_r}) over $t$ gives 
\begin{equation}
\frac{\partial R}{\partial t}=-\bigg[ \frac{2GM_{r}}{r_{0}^{3}}\frac{(1-R)}{R}\bigg]^{1/2}, 
\label{eq_velocity}
\end{equation}
which describes the infall velocity before the maximum collapse, i.e. when $R=0$.
The solution of equation (\ref{eq_velocity}) can be written in parametric form as follows:
\begin{eqnarray}
R &=& \cos^{2}\beta \label{parametric_r} \\
t &=& \bigg( \frac{r_{0}^{3}}{2GM_{r}}\bigg)^{1/2}\bigg[ \beta +\frac{\sin 2\beta}{2}\bigg].
\label{parametric_t}
\end{eqnarray}
For the uniform case, all mass shells will collapse simultaneously into a singularity at 
the free-fall time, $t_{ff}$, given by
\begin{equation}
t_{ff} = \sqrt{\frac{3\pi}{32G\rho_{0}}}. \label{t_ff}
\end{equation}

Consider now a system of $N$ particles; the local density fluctuation arising from 
finite $N$ makes the free-fall time of each particle scattered around the mean value (\ref{t_ff}).
Consequently, the dispersion of free-fall time can be written by
\begin{equation}
\delta t_{ff} = -\frac{1}{2}\bigg(\frac{\delta\rho_{0}}{\rho_{0}}\bigg) t_{ff}
\label{delta_tff}
\end{equation}
where $\delta\rho_{0}$ is the density fluctuation.
Then, it can be shown that the dispersion of infall velocity at any $t$ can be
approximated by
\begin{equation}
\delta\dot{R} \sim  \frac{\delta t_{ff}}{t_{ff}} \times \frac{\partial R}{\partial t}. \label{delta_rdot}
\end{equation}
The variation of $\delta\dot{R}$ means that the particles having shorter free-fall times 
will have greater infall velocities. This implies that with microscopic density fluctuation
the gravitational collapse is no longer in perfect singularity at $t_{ff}$. 
The minimal size, $R_{min}$, is instead attained which can be scaled with $N$ as, following
the analysis of \citet{boily+athanassoula+kroupa_2002}, 
\begin{equation}
R_{min}\propto N^{-1/3}. \label{r_min}
\end{equation}

Further from that the finite-$N$ effect prevents the collapse to attain the 
absolute singularity, the velocity dispersion (\ref{delta_rdot}) also becomes the source of 
the departure from spherical symmetry in the following manner. 
Given the particles falling to the origin from all directions with velocity 
dispersion in equation (\ref{delta_rdot}), they will be moving outwards with different velocities.
We speculate that during the process of shape evolution the particles with greater outward 
velocities tend to expand further and then they are more likely to establish the longer axis.
Before we start the estimation, we first 
suppose that, when the violent relaxation is finished after the cold collapse,
the final shape of bound structure is always in ellipsoidal form having three calculable 
semi-principal axes of lengths $a, b$ and $c$ such that $a \geq b \geq c$. To evaluate the 
degree of departure from spherical symmetry, we define the flattening $\iota$ as
\begin{equation}
\iota = \frac{a-c}{a}. \label{eq_flatness}
\end{equation}
By definition $\iota$ ranges from $0$ (sphere) to $1$ (thin disc). 
As it was noticed by \citet{casertano+hut_1985} that the determination of system size 
is significantly sensitive to the choice of the positioning of coordinate frame,
we then impose that the calculations of all physical quantities of the stationary  
structures are fixed on the centre of gravity coordinate.
This is also in correspondence with analysis of numerical results in Section \ref{results}.

To estimate the flattening of the final shape, let us suppose that the permitted 
relaxation time to the final shape is given by $t_{relax}$. We then estimate
the semi-major axis length by the distance spanned by a particle with maximum outward velocity, i.e.
\begin{equation}
a \sim  \dot{R}_{max} t_{relax}
\label{estim_a}
\end{equation}
where $\dot{R}_{max}$ is the maximum infall velocity. 
For the same reason, the semi-minor axis length corresponds to
\begin{equation}
c \sim  (\dot{R}_{max}-\delta \dot{R}) t_{relax}.
\label{estim_c}
\end{equation}
The expressions above imply that the key development to triaxiality relies on the 
dispersion of infall velocity passing at the origin at $t_{ff}$.
Then, we approximate the maximum velocity by
\begin{equation}
\dot{R}_{max} \sim \frac{R(0)}{t_{ff}-\delta t_{ff}}, 
\label{estim_rdotmax}
\end{equation}
where $R(0)$ is the initial position. If we suppose that $\delta t_{ff} \ll t_{ff}$,
we therefore obtain the estimate of $\iota$ of the stationary states to be
\begin{equation}
\iota \propto  \big<\delta\dot{R}\big>_{t=t_{ff}} \propto
\frac{\langle \delta t_{ff}\rangle }{t_{ff}} \bigg[ \frac{\partial R}{\partial t}\bigg]_{t=t_{ff}},
\label{estim_iota}
\end{equation}
where $\big< \ \big>$ denotes the average. To explain the approximation (\ref{estim_iota}) 
in other words, the origin of flattening comes from the combination of two 
main factors which are the dispersion of $t_{ff}$ and the mean infall velocity at $t_{ff}$.
Replacing the relations (\ref{eq_velocity}) and (\ref{delta_tff}) into  
(\ref{estim_iota}) and taking $R_{min} \ll 1$, the estimate becomes 
\begin{equation}
\iota \propto \bigg< \frac{\delta\rho_{0}}{\rho_{0}}\bigg> \frac{1}{R_{min}^{1/2}}.
\label{estim_iota2}
\end{equation} 
In the case of uniform density discretized into point-mass system of randomly placed 
$N$ particles, the average density fluctuation can be approximated by Poissonian noise, 
i.e. $\langle\delta\rho_{0}/\rho_{0}\rangle\sim 1/\sqrt{N}$. Following the 
scaling of $R_{min}$ given by equation (\ref{r_min}), the estimate of $\iota$ becomes
\begin{equation}
\iota\propto N^{-1/3}. \label{iota_n}
\end{equation}
The scaling (\ref{iota_n}) implies that the stationary state is less deformed as $N$ 
increases since the density fluctuation triggering the shape evolution is diminished. 
Note that the advantage of choosing the uniform case is that the density fluctuation is
purely that of Poisson, which is essentially a function of only $N$. Therefore, the 
estimated flattening of stationary states can be characterized just by a single parameter.
Analysis of non-uniform initial condition is more complicated as the 
density fluctuation does not only come from Poissonian noise. So the prediction might
involve more than one parameter.

In general, the violent relaxation of cold initial state is always followed by the
process of mass ejection that expels an amount of mass away from the system (see, e.g.,
\citealp{mcglynn_1984, aarseth+lin+papaloizou_1988, joyce+marcos+sylos_labini_2009, sylos_labini_2012, sylos_labini_2013}). 
Therefore, the mass of the virialized state is always reduced from that 
of initial state. In our analysis, we neglect all ejected particles and 
only the remaining bound part is in consideration.

\subsection{Generation of angular momentum} \label{theory_angular_momentum}
We investigate further the generation of angular momentum from cold collapse. 
Given the local density fluctuation, the trajectories of infall particles are meant 
to deviate from completely radial. Thus, at $t_{ff}$ when the particles pass by the 
origin, they can gain some rotational component. In cold non-rotating initial state,
we speculate that the angular momentum of relaxed structure is produced by the exchange 
between the escaping particles and the remaining bound particles during the violent relaxation.
The mass ejection is supposed to be asymmetric so that the exchanged 
angular momentum is not cancelled out.

From equation (\ref{eq_velocity}), we can deduce that at $t_{ff}$ the infall velocity
diverges in continuum limit. But if $N$ is finite, the same discreteness that prevents the 
singularity also imposes the limit of the infall velocity. 
We estimate the average infall velocity that can be attained at $t_{ff}$ 
by taking the limit $R \rightarrow R_{min}$ where $R_{min} \ll 1$ and we therefore have
\begin{equation}
v_{ff} = \sqrt{\frac{GM}{R_{min}}} \propto N^{1/6}. \label{v_max}
\end{equation}
Note that this expression is in line with \citet{joyce+marcos+sylos_labini_2009}. 
Suppose that the perpendicular distance to the origin of particles is given by 
$R_{min}$, the estimate of accumulated angular momentum 
of a system composed of $N_{b}$ bound particles of mass $m$ therefore reads
\begin{equation}
L \propto mN_{b}v_{ff}R_{min}. \label{l_total}
\end{equation}
To avoid the complexity arising from $N$-dependence of $N_{b}$, 
it is more convenient to consider the specific angular momentum $j=L/mN_{b}$.  
We therefore obtain the scaling of $j$ as a function of $N$ 
following the relations (\ref{r_min}) and (\ref{v_max}) as
\begin{equation}
j \propto v_{ff}R_{min} \propto N^{-1/6}. \label{spe_angu}
\end{equation}
This means that the net rotational motion vanishes in the continuum limit since the 
evolution is completely radial at all time.


\section{$N$-body simulation setup} \label{nume}
\subsection{Initial condition and units} \label{ic}
In this study, the initial condition is the sphere of uniform density composed of $N$
particles of identical mass where all particles are at rest (i.e. cold uniform sphere). 
To generate it for any $N$, particles are 
thrown randomly in a sphere of radius $r_{0}$ with uniform probability density, i.e.
\begin{equation}
P(\vec{r}) = \left\{ \begin{array}{lcl}
1/(\frac{4}{3}\pi r_{0}^{3}) & ; & |\vec{r}| \leq r_{0} \\
0 & ; & \textrm{elsewhere.}
\end{array}\right.
\end{equation}
The total mass $M$ is fixed to $\frac{4}{3}\pi r_{0}^{3}$; thus, the mass of 
individual particle $m$ is equal to $\frac{4}{3}\pi r_{0}^{3}/N$. Note that by 
this choice of parameters, the initial density $\rho_{0}$ is always equal to $1$.
The initial potential energy of a sphere is thus
\begin{equation}
U_{0} = -\frac{3GM^{2}}{5r_{0}}.
\end{equation}
Here we put $r_{0}=0.5$ for all simulations. 
For temporal evolution, the time unit is presented in the free-fall time given by
\begin{equation}
t_{dyn}=\sqrt{\frac{3\pi}{32G\rho_{0}}}.
\end{equation}
In statistical point of view, with $G$, $M$ and $r_{0}$ fixed, 
it remains only the random microscopic density fluctuation to vary in each realization
with mean density contrast decaying with $N$. To put it simply, if we fix those 
three parameters, there is only a single parameter that characterizes
the initial condition. For convenience we choose it to be $N$.
This choice is proper when comparing with the theoretical predictions as we 
derive them in terms of $N$.


\subsection{Simulation parameters} \label{setup}
Equation of motion of each particle is integrated using GADGET-2 code in public version
(see \citealp{springel+yoshida+white_2001, springel_2005} for guidance). We consider 
purely self-gravitating system in non-expanding background. 
The gravitational force is softened by a spline 
with adjustable softening length $\varepsilon$ in the
way that it converges to Newtonian above $\varepsilon$, but below that length the force is 
decreasing and vanishes as distance goes to zero
(see \citealp{springel+yoshida+white_2001} for detail). 
Given an initial system size of order
$1$, the average inter-particle distance $\ell$ can be approximated by $\ell\approx 1/N^{1/3}$. 
We then adjust the softening length as decreasing function of $N$ as
\begin{equation}
\varepsilon = 0.028\ell = \frac{0.028}{N^{1/3}}. 
\end{equation}
We choose $\varepsilon$ much smaller than $\ell$ to ensure that the Newtonian gravity 
is well preserved in close encounters during the collapse. 
Although it was found by \cite{boily+athanassoula_2006} that varying $\varepsilon$ has 
significant change on the final shapes, our choice of $\varepsilon$ is well below the 
range where the final shapes are sensitive to it.
We can hereby neglect the dependence on $\varepsilon$ in our study.

In dynamical integration, the total force on an individual particle 
is determined by direct summation at each time step. 
We separate the simulation into three phases: pre-collapse ($0-0.5 \ t_{dyn}$), 
collapse ($0.5-1.5 \ t_{dyn}$) and post-collapse (after $1.5 \ t_{dyn}$). 
During each phase, the integration time step is fixed.
The time steps for pre-collapse and collapse, where the maximum contraction occurs, 
are controlled to be around $t_{dyn}/2000$ and $t_{dyn}/20000$, respectively. 
In post-collapse, it varies from $t_{dyn}/5000$ to $t_{dyn}/10000$ so that,
in all realizations, the error on total energy is less than $0.2 \%$ 
for $N < 10000$ and $0.5 \%$ for $N > 10000$ at the end of simulations.


\section{Numerical results} \label{results}

The numerical simulations are performed for $N=1000$, $2000$, $4000$, $8000$, $16000$ and 
$32000$ with $70$, $40$, $30$, $30$, $6$ and $2$ different realizations, respectively.
All simulations are terminated at $t=14.28 \ t_{dyn}$ at first, with prolongation to 
$28.56 \ t_{dyn}$ for a few selected cases with $N=16000$ and $32000$ for specific inquiry. 
Here, the system of $N$ which is not too large is preferred in order to preserve the 
statistical effect from density fluctuation that diminishes substantially with $N$.
We overcome the anticipated uncertainty arising in small $N$ system by the 
ensemble average over considerable number of realizations if necessary.

\subsection{Temporal evolution of $\iota$} \label{evolution_shape}

\begin{figure*}
\begin{tabular}{ccc}
  \includegraphics[width=5.5cm]{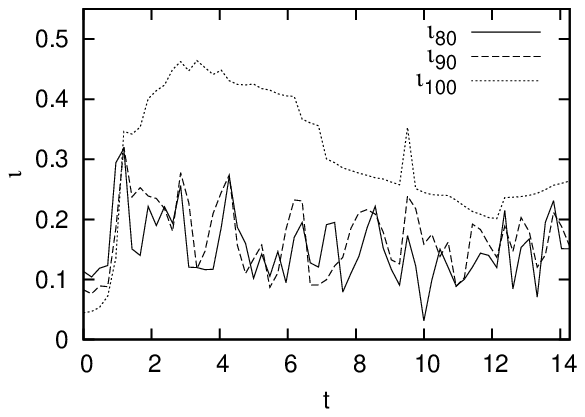} & \includegraphics[width=5.5cm]{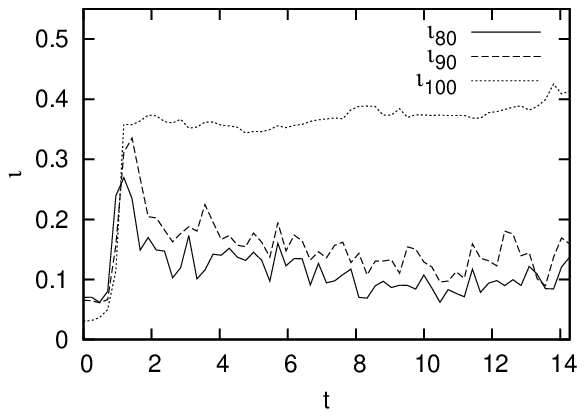} & \includegraphics[width=5.5cm]{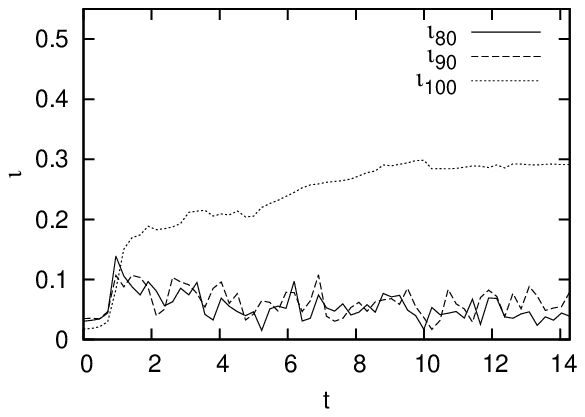} 
\end{tabular}
\caption{Temporal evolutions of $\iota_{80}, \iota_{90}$ and $\iota_{100}$ for selected 
realizations with $N=1000$ (left), $4000$ (middle) and $8000$ (right). 
Time unit is presented in dynamical time.}
\label{fig_i8910}
\end{figure*}

\begin{figure*}
\begin{tabular}{ccc}
\hspace{-13mm} \includegraphics[width=7cm]{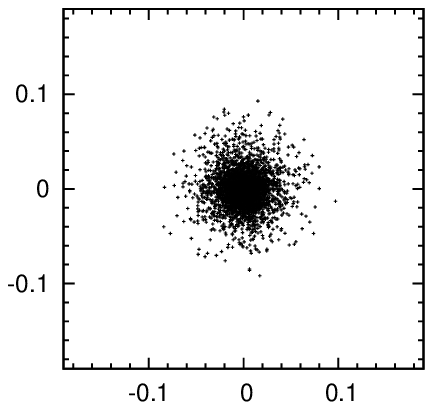} \hspace{-13mm} & 
\includegraphics[width=7cm]{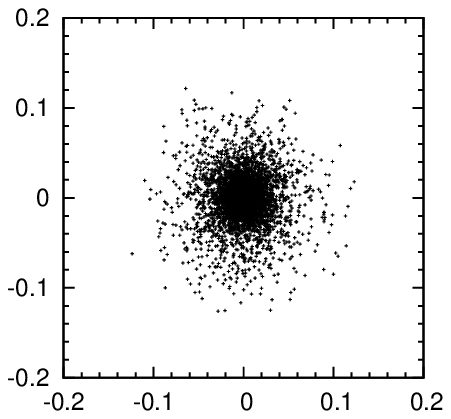} \hspace{-13mm} & 
\includegraphics[width=7cm]{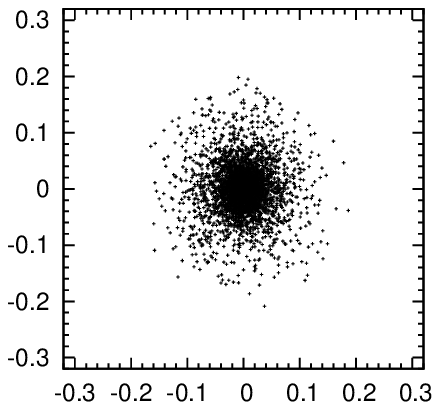} \\ 
\hspace{-13mm} \includegraphics[width=7cm]{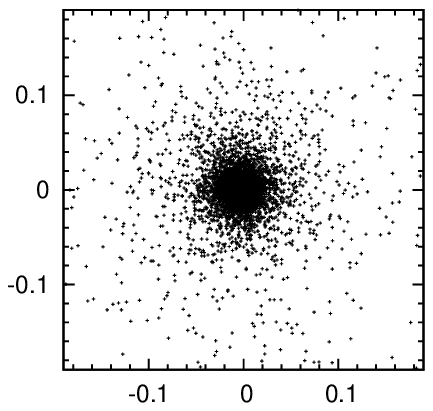} \hspace{-13mm} & 
\includegraphics[width=7cm]{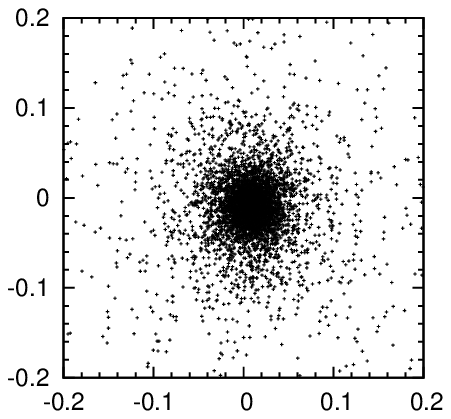} \hspace{-13mm} & 
\includegraphics[width=7cm]{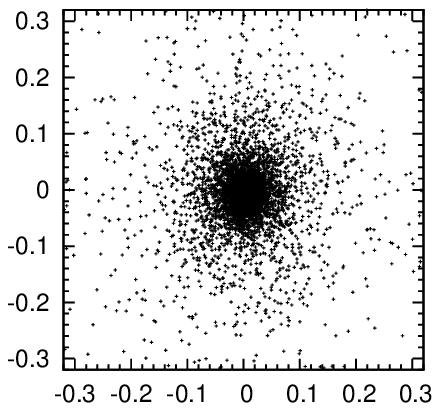} 
\end{tabular}
\caption{Configurations of three selected systems with $N=8000$ at the end of simulations. 
Two left-hand panels correspond to the case in Fig. \ref{fig_i8910} (right) with 
$\iota_{80}=0.039$. Two middle panels and 
two right-hand panels show two cases with $\iota_{80}=0.132$ and $0.142$, respectively, which are
the two most flattened cases. Three top panels indicate distributions of $80\%$ 
most bound mass projected on to their $(\hat{e}_{zz},\hat{e}_{xx})$ planes while three bottom 
panels correspond to distributions of $100\%$ bound mass projected on to the same 
planes of their $80\%$ bound structures. Length unit is defined in Section \ref{ic}.}
\label{fig_80100}
\end{figure*}

\begin{figure*}
\begin{tabular}{ccc}
\hspace{-13mm} \includegraphics[width=7cm]{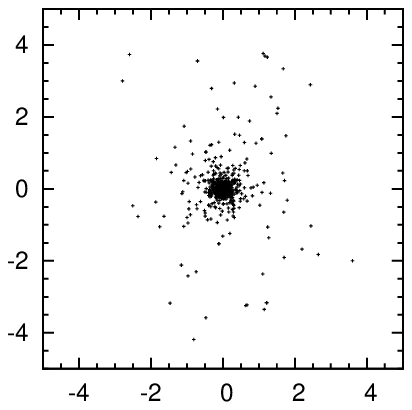} \hspace{-13mm} & 
\includegraphics[width=7cm]{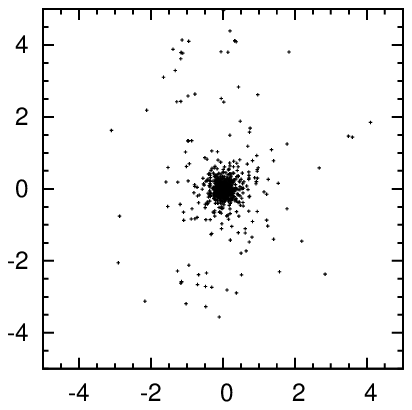} \hspace{-13mm} & 
\includegraphics[width=7cm]{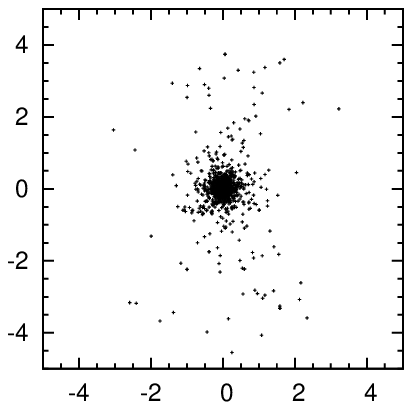}
\end{tabular}
\caption{Distributions of $100\%$ bound mass projected on to their $(\hat{e}_{zz},\hat{e}_{xx})$ planes
for the same three cases presented in Figs. \ref{fig_80100} in the enlarged coordinate frame.}
\label{fig_100}
\end{figure*}

We consider first the temporal evolutions of shapes of the bound structures.
In correspondence with the theoretical framework, 
the stationary states formed by bound particles are assumed to be ellipsoidal. 
To determine the semi-principal axis lengths in the course of evolution, we calculate 
first the moment of inertia tensor at any time $t$
\begin{equation}
I(t) = 
\begin{bmatrix}
I_{x'x'}(t) & I_{x'y'}(t) & I_{x'z'}(t) \\
I_{y'x'}(t) & I_{y'y'}(t) & I_{y'z'}(t) \\
I_{z'x'}(t) & I_{z'y'}(t) & I_{z'z'}(t)
\end{bmatrix}
\label{i_inertia}
\end{equation}
where each element is determined in the centre of mass coordinate 
of the bound system. The moments of inertia about three principal axes, 
i.e. $I_{xx}$, $I_{yy}$ and $I_{zz}$, correspond to the eigenvalues of the 
matrix (\ref{i_inertia}). We arrange them such that $I_{xx} \leq I_{yy} \leq I_{zz}$
and then the lengths of semi-principal axes associated 
with $I_{xx}$, $I_{yy}$ and $I_{zz}$ are, respectively,
\begin{eqnarray}
a &=& \sqrt{\frac{5}{2mN_{b}}(I_{yy}+I_{zz}-I_{xx})} \nonumber \\
b &=& \sqrt{\frac{5}{2mN_{b}}(I_{zz}+I_{xx}-I_{yy})} \label{abc_ixyz} \\
c &=& \sqrt{\frac{5}{2mN_{b}}(I_{xx}+I_{yy}-I_{zz})}. \nonumber
\end{eqnarray}
By our manipulation we then have $a \geq b \geq c$. The eigenvectors corresponding to those 
three principal axes, i.e. $\hat{e}_{xx}$, $\hat{e}_{yy}$ and $\hat{e}_{zz}$, 
are also calculated for further purpose. 

Here, we consider $\iota$ of three different fractions of mass which are $80\%$, $90\%$ 
most bound mass and $100\%$ bound mass, hereby $\iota_{80}$, $\iota_{90}$ and $\iota_{100}$, respectively.
Shown in Fig. \ref{fig_i8910} are temporal evolutions of $\iota_{80}$, $\iota_{90}$ and 
$\iota_{100}$ for selected realizations with $N=1000, 4000$ and $8000$. 
Consider first $\iota_{80}$ and $\iota_{90}$; they increase and settle down to 
the stationary states giving non-zero flattenings within a few oscillations. 
This stage corresponds to the collisionless (or mean-field) relaxation as demonstrated,
using the same type of initial condition, by \citet{joyce+marcos+sylos_labini_2009}. 
It was shown by them that the virialization was fully achieved by this process.
The evolutionary paths of both parameters are qualitatively similar.
The fluctuation of the stationary state is observed, which is more smoothed out as $N$ increases.
This persistence of oscillations with period about the dynamical time
may correspond to the remnant of the 
parametric resonance, originally studied by \citet{levin+pakter+rizzato_2008}, 
which is specific for the violent relaxation of cold initial condition and is responsible
for the mass and energy ejection.
In contrast, $\iota_{100}$ evolves differently as it separates from $\iota_{80}$ 
and $\iota_{90}$ since the first dynamical time.
Its evolution is moreover unpredictable as, after the cold collapse,
it can increase, decrease or remain unchanged before reaching the stationary value. 
The relaxation times of $\iota_{100}$ vary considerably from realization to realization.
The flattening estimated from $\iota_{100}$ is higher than using $\iota_{80}$ and $\iota_{90}$ but 
the fluctuation is more smoothed out and the amplitude is apparently uncorrelated to $N$.

We investigate further the spatial configurations of the stationary states.
Shown in Figs. \ref{fig_80100} are the distributions of $80$ and $100$ per cent bound mass for 
three selected cases with $N=8000$ at $t=14.28 \ t_{dyn}$. One of them corresponds to the case 
in the right-hand panel of Fig. \ref{fig_i8910}, with $\iota_{80}=0.039$, and two others are with 
$\iota_{80}=0.143$ and $0.123$, representing the two most flattened realizations for this $N$. 
In each case, both snapshots are projected on to the $(\hat{e}_{zz},\hat{e}_{xx})$ plane of the $80$ per cent
bound mass, i.e. the plane where we see the structure of $80$ per cent bound mass most elongated in the vertical 
axis, with use of the eigenvectors determined beforehand. Considering first of all the configurations of 
$100$ per cent bound mass, we find that these structures are constituted from the central dense region 
with surrounding particles in the outer diluted region. This kind of structure can be 
identified as the 'core-halo' that was discovered in many previous works. 
When comparing these two fractions of mass, we find that the central regions of $100$ per cent bound 
structures are similar to the $80$ per cent bound structures while the difference between those two 
is more remarked in the surrounding region. 
So, it is reasonable to say that the $80$ per cent bound mass represents the core of the stationary state.
Now let us examine the configurations of the cores. We observe in all three cases the elliptical 
cores with various sizes and flattenings in line with the reported $\iota_{80}$. 

To investigate more in detail the $100$ per cent bound structures,
the distributions of $100$ per cent bound mass for the same cases in Fig. \ref{fig_80100} 
are plotted in Fig. \ref{fig_100}, now in enlarged coordinate frames and projected on to 
their own $(\hat{e}_{zz},\hat{e}_{xx})$ planes. 
In these figures, we find distant particles far from the core and most of them
are situated near the major principal axis. 
However, it is clear that the projected configurations when including those outermost particles 
in the halo are not elliptical. 
This explains why the flattening estimated from $\iota_{100}$ is higher than using 
$\iota_{80}$ or $\iota_{90}$ since those particles, albeit a few,
are dominant in calculating $\iota_{100}$.
It therefore turns out to be that the parameter $\iota_{100}$ can attain
the stationarity even though the entire configuration is not necessarily ellipsoidal.

These results suggest that $\iota_{80}$ is appropriate for studying the intrinsic shapes 
of stationary states but, in contrast, $\iota_{100}$ is found to be inaccurate. 
We also find no significant difference between using $\iota_{80}$ and $\iota_{90}$.


\subsection{$N$-dependence of $\iota$} \label{scaling}

\begin{figure}
\begin{tabular}{c}
 \hspace{-4mm} \includegraphics[width=9cm]{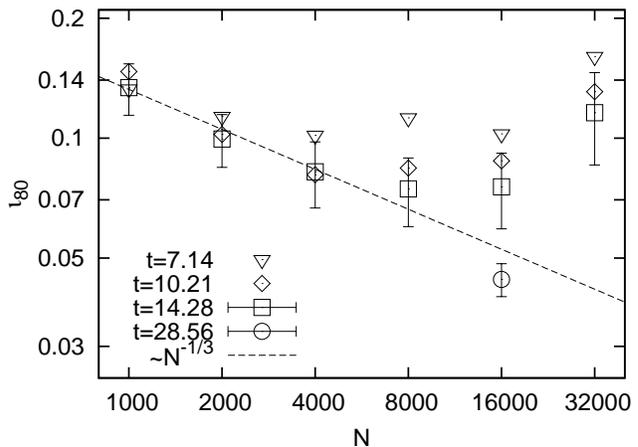}
\end{tabular}
\caption{Log-log plot of ensemble-averaged $\iota_{80}$ as a function of $N$ measured at
$t=7.14, 10.21$ and $14.28 \ t_{dyn}$, with a few cases extended to $28.56 \ t_{dyn}$. 
Dashed line corresponds to the $N^{1/3}$ best fit following equation (\ref{iota_n})
with data from $N < 10000$ at $t=14.28 \ t_{dyn}$. Widths of error bars  
are estimated from standard deviation over realizations, except for $N=32000$ where 
it corresponds to the difference between two realizations.}
\label{fig_scalingi80}
\end{figure}

\begin{figure*}
\begin{tabular}{cc}
  \includegraphics[width=9cm]{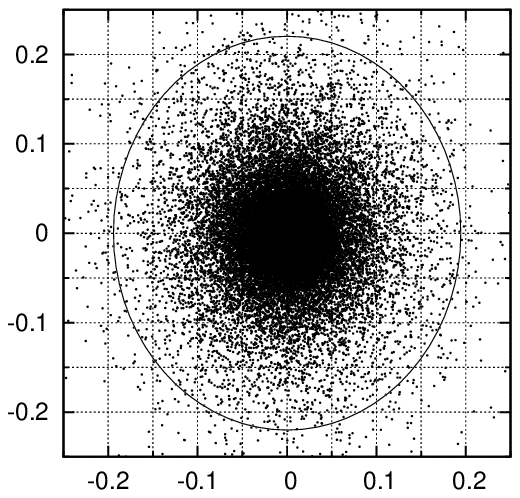} \hspace{-15mm} & \includegraphics[width=9cm]{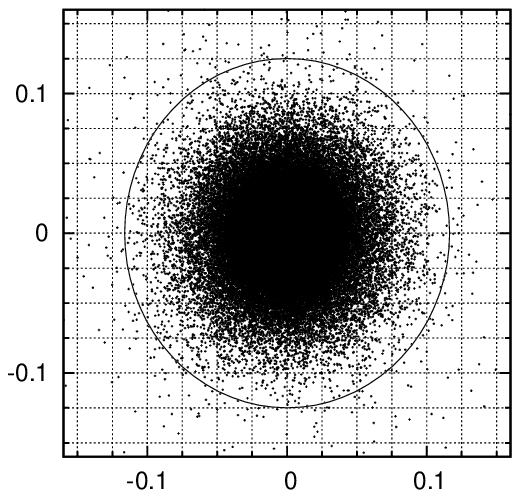} 
\end{tabular}
\caption{Superpositions of $(\hat{e}_{zz},\hat{e}_{xx})$ projections of $80\%$ bound particles
from all realizations with $N=1000$ (left) and $N=8000$ (right) at $t=14.28 \ t_{dyn}$. 
The ensemble-averaged $\iota_{80}$ are equal to $0.13$ and $0.075$, respectively. 
Ellipses in solid lines correspond to the $\iota_{80}$ at that time.}
\label{fig_stack}
\end{figure*}

In this section, we examine the $N$-dependence of the flattening of the stationary states. 
First, the log-log plot of ensemble-averaged $\iota_{80}$ from simulations as a function of $N$ 
at different time slices after virialization is shown in Fig. \ref{fig_scalingi80}.
To obtain each point, the flattening is determined individually in each realization
before being averaged.
Straight line corresponds to the $N^{-1/3}$ best fit following equation (\ref{iota_n})
with points from $N \leq 8000$ at $t=14.28 \ t_{dyn}$. 
Error bars at $t=14.28 \ t_{dyn}$ correspond to the standard deviation calculated over realizations, 
except for $N=32000$ where the width corresponds to the difference between two realizations.
Let us consider first the evolution of first three time slices. 
We see that the prediction fits well with simulated results from $N=1000$, $2000$ and $4000$ at $t=14.28 \ t_{dyn}$.
At $7.14 \ t_{dyn}$, it appears that the points from two latter cases are not completely relaxed, which 
might indicate another evolution stage that reshapes the virialized core to the prediction. 
This evolution is seen more clearly in the $N=8000$ case in which 
at $t=7.14 \ t_{dyn}$ the measured $\iota_{80}$ is above the $N^{-1/3}$ line while, as time progresses, 
it is approaching down to the prediction. At $t=14.28 \ t_{dyn}$, the point is acceptably close to the prediction. 
We also notice this evolution for $N=16000$ and $32000$  
but the simulations cannot attain the predicted flattenings within that time, 
with discrepancy growing with $N$. 
These tendencies may imply that the relaxation is still undergoing and the overall 
agreement may possibly be improved if the simulations are extended further.
To verify this, four selected cases for $N=16000$ are prolonged 
to $28.56 \ t_{dyn}$. The ensemble-averaged $\iota_{80}$ is included in the figure with 
error bar corresponding to the standard deviation of the extended realizations. 
We find that the point passes the prediction line and stays slightly below. At 
this time, it is closer to the prediction than the previous snapshot.

\begin{figure}
\begin{tabular}{c}
 \hspace{-4mm} \includegraphics[width=9cm]{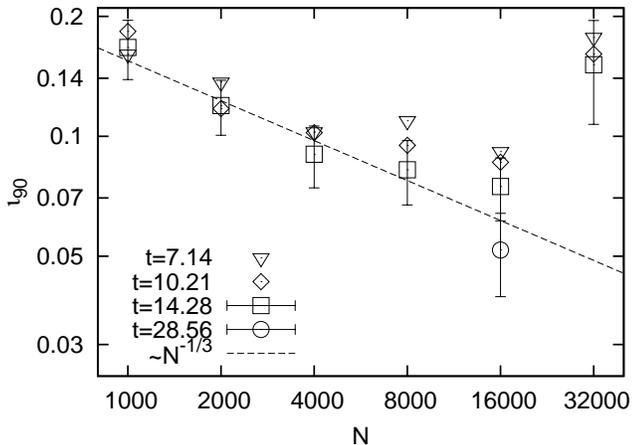}
\end{tabular}
\caption{Log-log plot of ensemble-averaged $\iota_{90}$ as a function of $N$ measured at
$t=7.14, 10.21$ and $14.28 \ t_{dyn}$, with a few cases extended to $28.56 \ t_{dyn}$. 
Dashed line corresponds to the $N^{1/3}$ best fit following equation (\ref{iota_n})
with data from $N < 10000$ at $t=14.28 \ t_{dyn}$. 
Error bars are calculated in the same way as in Fig. \ref{fig_scalingi80}.}
\label{fig_scalingi90}
\end{figure}

\begin{figure*}
\begin{tabular}{cc}
 \hspace{-8mm} \includegraphics[width=8.5cm]{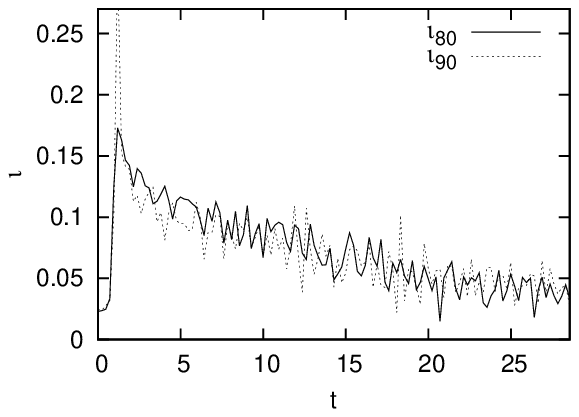} & \includegraphics[width=8.5cm]{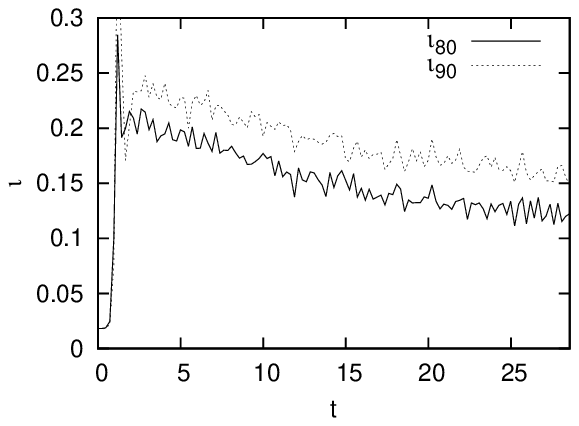} 
\end{tabular}
\caption{Extended temporal plots of $\iota_{80}$ and $\iota_{90}$ for selected 
realizations with $N=16000$ (left) and $32000$ (right).}
\label{fig_i1632}
\end{figure*}

To verify if the ensemble-averaged $\iota_{80}$ is in coherence with the real flattening of stationary state,
the superpositions of $(\hat{e}_{zz},\hat{e}_{xx})$ projections at $t=14.28 \ t_{dyn}$ for $80\%$ 
bound particles from all realizations with $N=1000$ and $8000$ are shown in Fig. \ref{fig_stack}.
Projection of each realization is produced separately on to its $(\hat{e}_{zz},\hat{e}_{xx})$ 
plane before being superposed. Ellipses in solid lines corresponding to
$\iota_{80}$ at that time are put for comparison.
We find that both configurations are in good forms with the ellipses. 
It therefore appears that $\iota_{80}$ is an appropriate measure to represent
the flattening of the stationary state.

\begin{figure*}
\begin{tabular}{cc}
  \includegraphics[width=9cm]{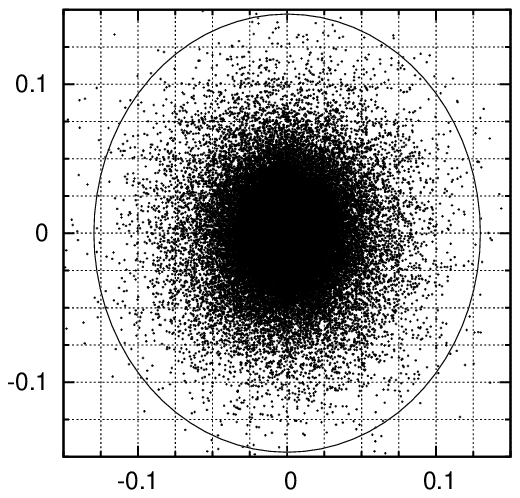} \hspace{-15mm} & \includegraphics[width=9cm]{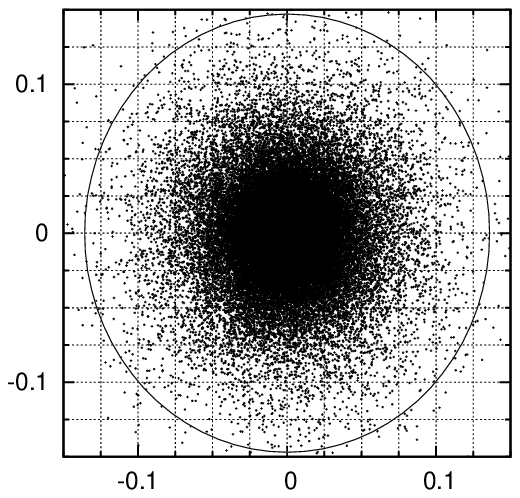} 
\end{tabular}
\caption{Superpositions of $(\hat{e}_{zz},\hat{e}_{xx})$ projections of $80\%$ bound particles 
from all realizations with $N=4000$ at $t=3.57 \ t_{dyn}$ (left) and $t=14.28 \ t_{dyn}$ (right). 
The ensemble-averaged $\iota_{80}$ are equal to $0.13$ and $0.082$, respectively. 
Ellipses in solid lines correspond to $\iota_{80}$ at that time.}
\label{fig_stack2}
\end{figure*}

Then the plot of ensemble-averaged $\iota_{90}$ as a function of $N$ for the same time 
slices is shown in 
Fig. \ref{fig_scalingi90} with the $N^{-1/3}$ best fit with points from $N \leq 8000$ cases 
at $t=14.28 \ t_{dyn}$ in dashed line. The error bars are determined in 
the same way as described above. Comparing with the plot of $\iota_{80}$ in 
Fig. \ref{fig_scalingi80}, we find that the flattening derived from $\iota_{90}$ is 
slightly greater. The overall appearance is similar and the $N^{-1/3}$ decay is 
still retained. We also note the late evolution towards the prediction after the 
virialization for $N=16000$.

From these two plots, we find an interesting remark that both $\iota_{80}$ and $\iota_{90}$ 
for $N=16000$ require relaxation time to prediction longer than $14.28 \ t_{dyn}$, while for 
smaller $N$ they are relaxed to the prediction well before with fluctuations around the 
predicted stationary states observed. To inspect this evolution, the extended plots of 
temporal evolutions of $\iota_{80}$ and $\iota_{90}$ with $N=16000$ and $32000$,
one realization for each $N$, are shown
in Fig. \ref{fig_i1632}. In these figures, we capture the slow decay of both $\iota$.
For $N=16000$ the attainment to stationary values is observed at $t \sim 20 \ t_{dyn}$.
Thus, we can safely presume the stationarity at $t=28.56 \ t_{dyn}$ for $N=16000$ in 
Figs. \ref{fig_scalingi80} and \ref{fig_scalingi90}. 
For $N=32000$, the decay lasts longer. We observe the apparent stationarity of $\iota_{80}$ 
at $t\sim 25 \ t_{dyn}$ but it is not clear for $\iota_{90}$. 
From these results, we verify the existence of secondary evolution that leads the virialized core 
to the state with lower $\iota$, but still non-zero.
To inspect further in configuration space,
the superpositions of $(\hat{e}_{zz},\hat{e}_{xx})$ projections for $80\%$ bound mass
with $N=4000$ at $t=3.57 \ t_{dyn}$ and $14.28 \ t_{dyn}$ are shown in 
Fig. \ref{fig_stack2}, with ellipses equivalent to $\iota_{80}$ at that time as references. 
Note that the former is taken just after the virialization with $\iota_{80}=0.13$ and the 
latter is when the system is already in the predicted stationary state with $\iota_{80}=0.082$.
We find that the projected core is relaxed to elliptical form since shortly after 
the virialization and, following the reported $\iota_{80}$, 
it is more rounded out at the end of simulation. 
Further discussion concerning these issues 
with reference to past study will be provided in Section \ref{conclusion}.

\begin{figure}
\begin{tabular}{c}
 \hspace{-4mm} \includegraphics[width=9cm]{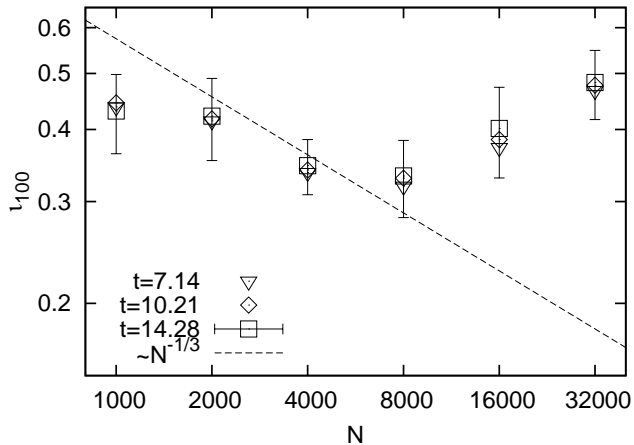}
\end{tabular}
\caption{Log-log plot of ensemble-averaged $\iota_{100}$ as a function of $N$ measured at
$t=7.14, 10.21$ and $14.28 \ t_{dyn}$.
Dashed line corresponds to the $N^{-1/3}$ best fit from equation (\ref{iota_n}) performed with data
at $t=14.28 \ t_{dyn}$. 
Error bars are calculated in the same way as in Fig. \ref{fig_scalingi80}.}
\label{fig_i100}
\end{figure}

Now the ensemble-averaged $\iota_{100}$ as a function of $N$ with error bars is shown in 
Fig. \ref{fig_i100}, along with $N^{-1/3}$ the best fit with points at $t=14.28 \ t_{dyn}$ for
comparison. We see that the plot of $\iota_{100}$ is different from two previous $\iota$ 
in many aspects. The flattening estimated from $\iota_{100}$ is, as expected, significantly 
greater but there is relatively low dispersion of $\iota_{100}$ from three time slices. 
The measured $\iota_{100}$ clearly does not follow the $N^{-1/3}$ prediction. 
The $N$-dependence is even absent in this plot. 
Unlike two previous $\iota$, it seems 
that extending the simulations will not change the results any further.
The fact that $\iota_{100}$ behaves differently from $\iota_{80}$ and $\iota_{90}$
could be explained by, in continuity from Section \ref{evolution_shape}, 
that it is strongly correlated to the outermost halo.
This part is shown to be decoupled from the core since the very beginning of 
the violent relaxation and makes the shape determination improper.

In summary, this section demonstrates clearly the $N$-dependence of the cores in 
the way that it tends to be more flattened out as $N$ decreases. The relaxed cores are in 
good agreement with the prediction provided in Section \ref{theory_evolution}.
For large $N$ systems, it appears that the relaxation process is longer, which reflects
the secondary evolution after the virialization. This process can be seen by the slow 
evolution towards more spherical, but still triaxial, configuration.


\subsection{Variation of intrinsic three-dimensional shapes} \label{pro_ob}

We study in this section how the final intrinsic shapes 
vary in different realizations. Note that we anticipate the variation of shape
because of the presence of the microscopic density fluctuation that is distributed randomly
in each realization. Given a system of semi-principal 
axes of lengths $a$, $b$ and $c$, we define the triaxial indices
\begin{equation}
\alpha = \frac{a-b}{b} \ \ \textrm{and} \ \ \gamma = \frac{b-c}{b}.
\end{equation}
We suppose that only four forms of ellipsoids are attainable: prolate, oblate, triaxial and 
spherical. Then the criterion to allocate to these four shapes is as follows:
\begin{equation}
\begin{array}{lclcl}
    \textrm{prolate} & : & \alpha > \delta & \textrm{and} & \gamma \leq \delta \\
    \textrm{oblate} & : & \alpha \leq \delta & \textrm{and} & \gamma > \delta \\
    \textrm{triaxial} & : & \alpha > \delta & \textrm{and} & \gamma > \delta \\
    \textrm{spherical} & : & \alpha \leq \delta & \textrm{and} & \gamma \leq \delta 
\end{array} \label{criteria_shape}
\end{equation}
where $\delta$ is adjustable.

\begin{figure*}
\begin{tabular}{cc}
 \hspace{-8mm} \includegraphics[width=9.5cm]{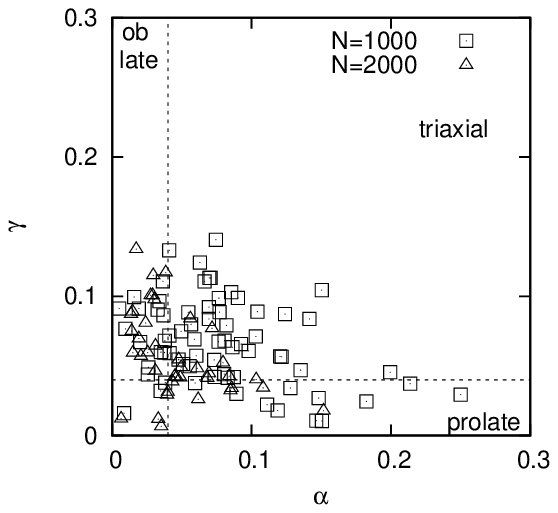} \hspace{-18mm} & \includegraphics[width=9.5cm]{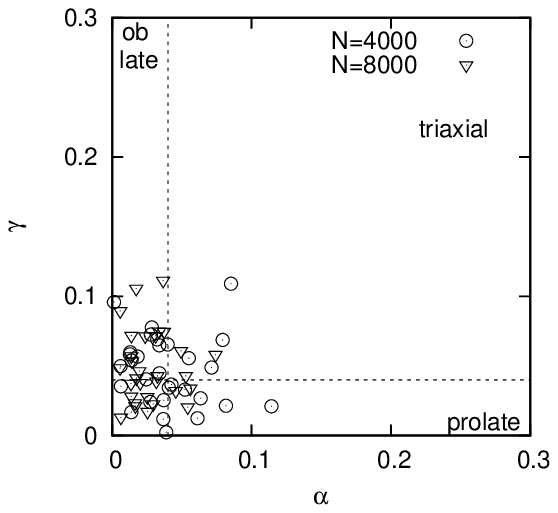} 
\end{tabular}
\caption{Plots of $(\alpha ,\gamma )$ from all realizations calculated from $80\%$ bound mass for 
$N=1000$, $2000$ (left), $4000$ and $8000$ (right) at $t=14.28 \ t_{dyn}$.
Separating lines following the criterion (\ref{criteria_shape}) with $\delta =0.04$ 
are put in dashed.}
\label{fig_pro_ob_80}
\end{figure*}

\begin{figure*}
\begin{tabular}{cc}
 \hspace{-8mm} \includegraphics[width=9.5cm]{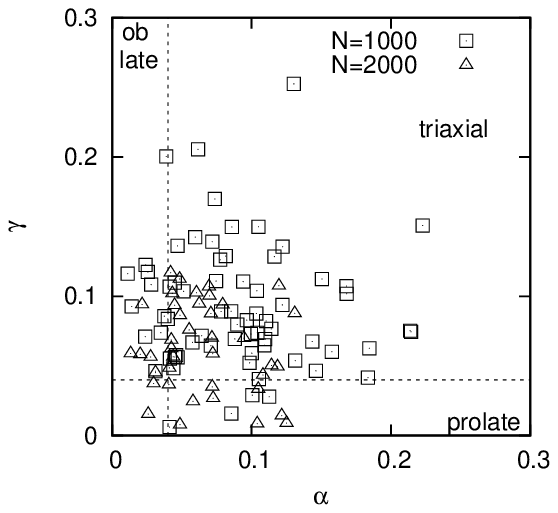} \hspace{-18mm} & \includegraphics[width=9.5cm]{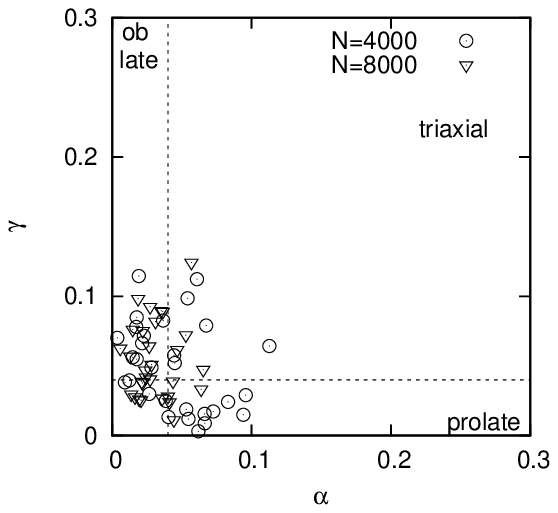} 
\end{tabular}
\caption{Plots of $(\alpha ,\gamma )$ from all realizations calculated from $90\%$ bound mass for 
$N=1000$, $2000$ (left), $4000$ and $8000$ (right) at $t=14.28 \ t_{dyn}$.
Separating lines following the criterion (\ref{criteria_shape}) with $\delta =0.04$ 
are put in dashed.}
\label{fig_pro_ob_90}
\end{figure*}

The intrinsic shapes of all realizations at $t=14.28 \ t_{dyn}$ for $N$ equal 
to $1000$, $2000$, $4000$ and $8000$
are summarized in $(\alpha ,\gamma )$ plots in Figs \ref{fig_pro_ob_80} and 
\ref{fig_pro_ob_90} for $80$ and $90$ per cent bound mass, respectively.
In this section, the cases with $N > 8000$ are excluded from our study as 
it was seen that these systems may not be fully relaxed at that time.
Separating lines following the criterion (\ref{criteria_shape}) are drawn in dashed 
with $\delta=0.04$. 
According to the plots, the intrinsic shapes of all realizations are not unique as
they vary from nearly spherical, 
prolate, oblate to completely triaxial with various degree of triaxiality.
The points from $N=1000$ and $2000$ are spread out mostly in the prolate, oblate and triaxial 
regions but, when $N$ increases, the points are less scattered and 
tend to gather near the origin. This result thus demonstrates
once again the tendency towards spherical symmetry in system with large $N$.

\begin{figure*}
\begin{tabular}{cc}
 \hspace{-8mm} \includegraphics[width=8.5cm]{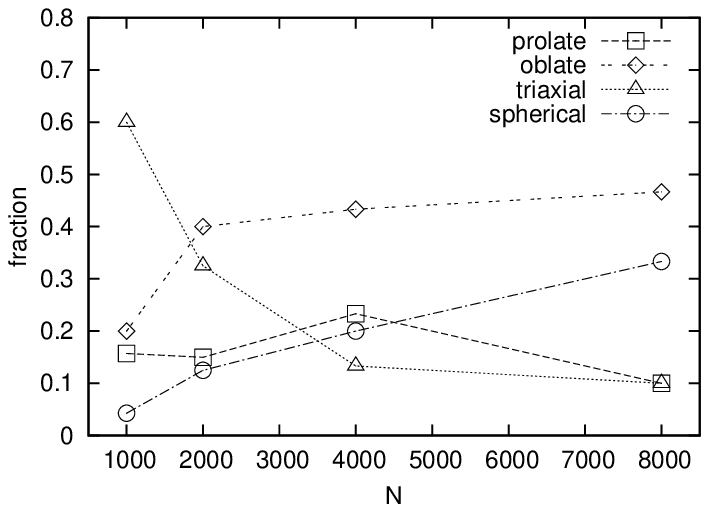} & \includegraphics[width=8.5cm]{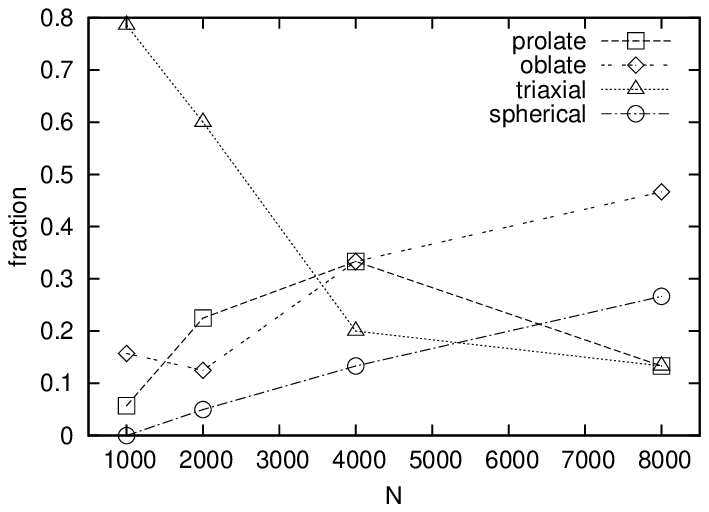} 
\end{tabular}
\caption{Plots of fractions of number of realizations in
prolate, oblate, triaxial and spherical shapes as a function $N$.
Left- and right-hand panels are for $80$ per cent bound mass and $90$ per cent bound mass, respectively.}
\label{fig_pro_ob_plot}
\end{figure*}

To study this topic more quantitatively, the fractions of number 
of realizations in each shape as a function of $N$ are plotted in Fig. \ref{fig_pro_ob_plot} 
for $80$ and $90$ per cent bound mass, using the same $\delta$ as in previous figures. 
Let us consider the $N$-dependence of each shape in both figures.
We find that the fraction of spherical increases with $N$,  
which emphasizes the convergence towards spherical symmetry as $N$ increases. 
The fraction of oblate ellipsoid also increases, on average, with $N$ while the fraction of 
triaxial decreases. For the prolate, the $N$-dependence is not simple. 
It increases at first and decreases when $N$ is greater than $4000$. 
At $N=1000$, the triaxial shape significantly dominates three others but as $N$ increases to $8000$   
the oblate becomes increasingly dominant.
When comparing between prolate and oblate, we find that the oblate ellipsoid is more favourable than
the prolate in the range of $N$ that we examine. 
The tendency towards oblateness in cold initial condition was also reported by 
\citet{boily+athanassoula_2006} but the opposite result supporting the prolate was 
also made by \cite{cannizzo+hollister_1992}. 
Thus, from those studies it appears that the conclusive answer for the preference between prolate 
and oblate is complicated as it correlates closely to the detail of initial density profile. 
However, these behaviours with $N$ we obtained in this section may not necessarily be the same 
if we consider the systems with much larger $N$. 
In that limit, we expect the spherical shape to be more dominant so the fractions of 
three other shapes should converge down to zero.

The related work by \citet{aguilar+merritt_1990} also studied the similar question but, 
differently from us, they were focusing rather on the influence from initial parameters 
such as virial ratio, axial ratio or velocity anisotropy to the final shapes. It was then revealed 
that the final shapes depended strongly on the choice of initial parameters.
Here, we investigate further the variation of final shapes when the initial condition 
is fixed to cold sphere of uniform density, leaving only the Poissonian density fluctuation 
in each realization to be varied. We find that the variation of density fluctuation, 
albeit in microscopic scale, can finally lead to the configurations with diverse degree of triaxiality.
When $N$ increases, the density fluctuation diminishes so the diversity of shape is less observed.
Although the more precise $N$-dependence for each shape is not obtained in this section,
the fact that the density fluctuation also takes part in establishing the final shape 
is crucial and not to be neglected.


\subsection{Specific angular momentum of ellipsoid} \label{angular}

\begin{figure*}
\begin{tabular}{cc}
  \includegraphics[width=8.5cm]{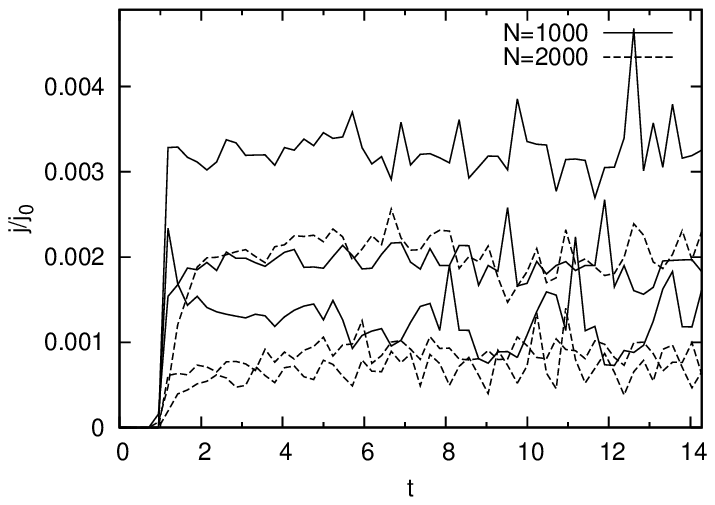} & \includegraphics[width=8.5cm]{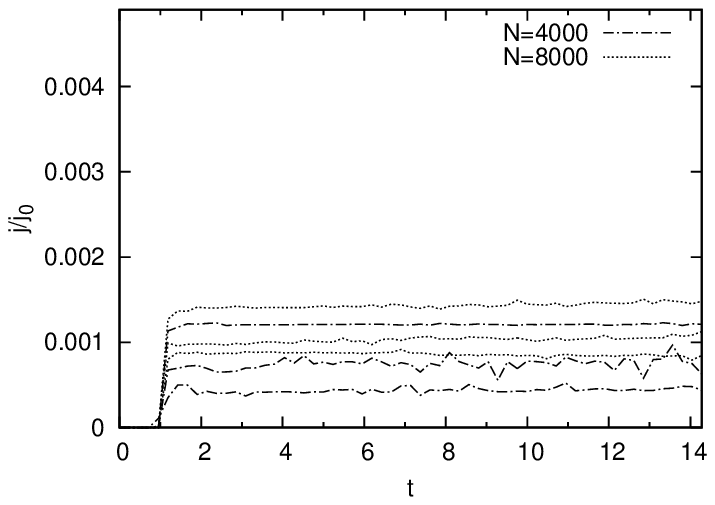} 
\end{tabular}
\caption{Temporal evolutions of dimensionless specific angular momentum, $j/j_{0}$, for 
$100$ per cent bound mass with $j_{0}$ given in the text. Left-hand panel is for the selected cases with
$N=1000$ and $2000$ while right-hand panel is for $N=4000$ and $8000$, three realizations 
for each $N$.}
\label{fig_l100t}
\end{figure*}

We present here the results about the specific angular momentum of bound structure. 
Let us recall that in the continuum limit the angular momentum is always absent 
since the motion is completely radial. While in the discrete system, the particles can have
mildly orbital motion due to the local density inhomogeneity.
The results in this section are arranged into two steps. First, we verify 
if systems possess the rotational motion after the mass ejection is finished.
Secondly, if the rotation is verified, we study further how it depends on $N$ in comparison with the
estimate that we provided in Section \ref{theory_angular_momentum}.

Plotted in Fig. \ref{fig_l100t} are temporal evolutions of dimensionless specific
angular momentum, $j/j_{0}$, computed from the $100$ per cent bound mass, and three selected cases 
each of $N=1000$, $2000$, $4000$ and $8000$ in the centre of mass frame.
The characteristic specific angular momentum $j_{0}$ is given by $j_{0}=r_{0}v_{0}$, 
where $v_{0} = \sqrt{\frac{GM}{r_{0}}}$. In the plots, we find the similarity of evolution 
pattern of $j/j_{0}$ in all presented cases regardless of $N$ which can be 
described as follows. First it is initially zero, as it should be in cold state, and the system 
persists to be rotationless until approximately the first dynamical time. Then $j/j_{0}$ 
increases sharply and relaxes shortly after, in concurrence with the virialization time. 
Thus from this result, we could be certain that the generation of angular momentum is initiated
by the mass ejection as we suspected before.
We find that the rotation is stable as, apart from the finite-$N$ fluctuation around the 
average, $j/j_{0}$ shows no evidence of other evolution to different state.
This might indicate that once the virialization is accomplished there is no further exchange 
of angular momentum between bound and ejected components. 
Also the angular momentum at stationary state varies considerably from realization to realization.

\begin{figure}
\begin{tabular}{c}
  \hspace{-4mm} \includegraphics[width=9cm]{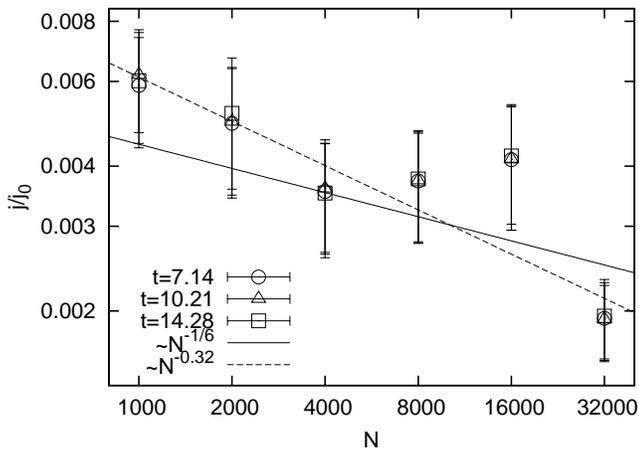}
\end{tabular}
\caption{Plot of ensemble-averaged $j/j_{0}$ as a function of $N$ measured at $t=7.14, 10.21$ and 
$14.28 \ t_{dyn}$. Solid line corresponds to the $N^{-1/6}$ best fit from equation (\ref{spe_angu}) with points 
from three snapshots. Dashed line is the $N^{-\lambda}$ best fit giving $\lambda =0.32$.  
Error bars are determined in the same way as for $\iota$ plots.}
\label{fig_l100}
\end{figure}

For more quantitative examination, the plot of 
ensemble-averaged $j/j_{0}$ with error bars as a function of $N$ from the $100$ per cent bound mass is shown 
in Fig. \ref{fig_l100}, taken at the same time slices as above. 
Error bars are determined in the similar way as for $\iota$ plots.
The $N^{-1/6}$ best fit from equation (\ref{spe_angu}) and the $N^{-\lambda}$ best fit giving 
$\lambda = 0.32$ are provided together. 
In this plot, both fittings are performed with data from three time slices.
We find that the dispersion of angular momentum from three snapshots is low
in accordance with the temporal plot in Fig. \ref{fig_l100t}. 
For the scaling of $j/j_{0}$, we find that the simulated $j/j_{0}$ 
matches better with $N^{-0.32}$, albeit with large fluctuation, than the $N^{-1/6}$ prediction.
From this result, the $N$-dependence of $j/j_{0}$ might reflect that the initial 
density fluctuation, which is the only $N$-dependent initial factor,
is also involved in the generation of angular momentum.
However, our explanation on this physical process is not yet satisfactory.
The question considering the origin of angular momentum is still under investigation.


\section{Conclusion and discussion} \label{conclusion}

In this paper, we study the problem of shape evolution in the $N$-body self-gravitating 
system starting from cold uniform sphere. 
Our main interest is on how well we can describe quantitatively 
the degree of deviation from spherical symmetry of the virialized structures. 
To begin with, we speculate that the origin of shape deformation from spherical symmetry 
is from the finite-$N$ density fluctuation in the way that the deformation
increases with the amplitude of that fluctuation.
Then by revisit of the dynamical model of spherical collapse, we implement the density 
fluctuation, which is purely Poissonian in our case, 
into the dynamical evolution and finally obtain the estimate of  
$\iota$ at stationary states as a function of $N$ to be $N^{-1/3}$.

In order to analyse the dynamics of shape evolution, the $N$-body simulations 
for various $N$ have been performed. The visual inspection of the 
stationary state in simulation suggests that the virialized state can be 
described by the so-called core-halo structure, as depicted by the 
central dense region (or core) loosely surrounded by the diffused region (or halo). 
It is subsequently shown that a sub-structure consisting of $80$ per cent most bound 
particles is acceptable to represent the core of the system and, likewise, 
the $90$ per cent most bound mass can also be used without any significant difference. 
For the three-dimensional configuration of the stationary states,
the temporal plots of $\iota$ reveal the departure from spherical symmetry 
of the cores during the violent relaxation, and the cores in virialized state can be 
approximated by ellipsoids with valid three semi-principal axis lengths.
Their intrinsic shapes exhibit various degrees of triaxiality and can be categorized 
into four different types: completely triaxial, prolate, oblate and nearly spherical.
In contrast, when considering the same questions for the entire 
core-halo system (i.e. the $100$ per cent bound mass), the problem is less 
clarified as the determination of shape is erroneous due to the disturbance
from distant particles in the outermost halo.

Focusing now on the $N$-dependence of the configurations, we find that the cores 
in simulations tend to be less flattened and their shapes are less diversified as $N$ increases. 
We find that the predicted $N^{-1/3}$ scaling of $\iota$ provides reasonably good 
agreement with the simulated cores in the range of $N$ from $1000$ to $16000$, with the extension 
of simulation required for $N=16000$. The increase of relaxation time with $N$ might manifest 
the intervention of collisional effect at this stage. This occurrence could be explained as follows. 
For the final state to deviate from the Vlasov limit, the scattering with the potential of 
neighbouring particles, which are distributed asymmetrically in microscopic scale, 
during close encounter is necessary.
In contrast, the related study led by \cite{roy+perez_2004} observed the opposite results 
that the $N$-dependence was 
missing for the cases with $N>30000$ in various initial conditions. Their conclusion may be 
true in some part as, from our studies on various system parameters for a considerable 
range of $N$, we find that the 
$N$-dependence behaviour of each parameter differs greatly. The parameter $\iota_{100}$ 
is apparently free from $N$ while the angular momentum exposes clearly the constant decreasing 
with $N$ for more than one order of magnitude (see more discussion below).
For $\iota_{80}$ and $\iota_{90}$, due to the increasing relaxation time with $N$, this question
is more difficult to resolve numerically for larger $N$.

When we look on the temporal plots of flattenings, they reveal obviously a distinguishable 
process of evolution to the prediction that lasts longer than the virialization. 
This secondary stage leads the virialized cores to more spherical, but still triaxial, states 
with relaxation time diverging with $N$. 
The sequential evolutions were also reported by \citet{theis+spurzem_1999}
where the authors demonstrated, using an appropriate measure of axial ratio, that 
shape evolution could be divided into three main stages: swift relaxation to triaxial 
configuration within a few $t_{dyn}$, readjustment to axisymmetric within 
$\sim 20 \ t_{dyn}$ and slow spherization that is continual until hundreds
of $t_{dyn}$. They asserted that the first two stages were in a collisionless regime with
time-scale independent of $N$
while only the third stage was collisional with time scale increasing with $N$. 
Back to our results, the two-step evolution scheme may coincide with their first two steps 
but we may have different opinions in two points. 
First, the underlying physical process of the second step should involve, either partly or fully, 
the collisionality. Thus, it appears that 
the collisional effect comes into regulating much earlier than previously thought by them.
Secondly the shape after the second step is not necessarily axisymmetric.
It was found that some final shapes ended up in triaxial.
However, because of the limited simulation time in our study, we are not able to verify 
the third step of evolution.

Shape deformation from spherical collapse in an isolated system 
has also been topic of research earlier by, for example, 
\citet{van_albada_1982, merritt+aguilar_1985, aguilar+merritt_1990, cannizzo+hollister_1992, theis+spurzem_1999, boily+athanassoula_2006} 
for systems with $N$ ranging from $10^{4}$ to $10^{6}$ with different types of initial conditions, 
mostly non-uniform. They found in some cases
that significant triaxiality emerged even if $N$ was as large as $8 \times 10^{5}$, while according
to our conclusion the triaxiality should not remain detectable in such a system. 
This unlikeliness can be explained by that they employed the inhomogeneous initial density 
profiles that yield the density contrast which does not decay with $N$ in addition to the 
Poissonian noise. 
Therefore, the dispersion of free-fall time is greatly enhanced even for very large $N$ systems.
Some simulations of cold collapse of power-law density profile with size comparable 
to our $N=8000$ system by \citet{cannizzo+hollister_1992} yielded considerably a more 
flattened structure. Their final axial ratio of $80\%$ bound mass was equivalent to 
$\iota$ between $0.28$ and $0.44$, depending strongly on the power-law 
index, which was more flattened than any realization in our study. 
However, it remains to be clarified if the same scaling is preserved for other type of 
initial condition. 
This macroscopic initial density contrast is interpreted as a `seed' of instability by
\cite{roy+perez_2004, marechal+perez_2010} which is compulsory in order to develop 
the highly elongated structures. On the contrary, they considered the homogeneous case 
as lacking this seed to progress effectively the shape deformation. 
From our simulations, we find that the stable triaxiality still can be developed in homogeneous 
initial state, though it is milder than the non-homogeneous one in past studies. 
Thus it turns out to be that the homogeneous case may carry `weak seed', which is of
course of microscopic origin and decays with $N$, rather than `no seed' as conjectured
by them.

In addition, we also examine the establishment of angular momentum of 
stationary state starting from non-rotating cold initial condition.
In past literature, the generation of angular momentum 
was mostly related to the tidal-torque theory (see \citealp{peebles_1969}) which
involved at least two systems exerting torque to each other.
Here, we propose alternatively that the rotation can also be generated in an
isolated system by angular momentum exchange between bound and ejected 
components during the asymmetric mass ejection.
This supposition is verified using numerical simulations that unveil the stable 
rotation and the $N$-dependence of angular momentum at stationary states.
The rotation of relaxed systems constantly slows down when $N$ increases.
We also note the short relaxation time of rotation and it is apparently independent of $N$.
The $N$-dependence imprinted in the angular momentum is the crucial point
which indicates that its origin is somehow related to
the initial density fluctuation. However, the underlying process is not yet understood. 

\vspace{0.2cm}

The author fully acknowledges the financial support granted by the Faculty of Science and
Technology of the Rajamangala University of Technology Suvarnabhumi as a crucial part
in conducting the research. Also useful discussions from Michael Joyce and 
Francesco Sylos Labini are grateful.

\balance

\bibliographystyle{mn2e}

\label{lastpage}

\end{document}